\documentclass[journal]{IEEEtran}
\usepackage{bbm}
\usepackage{mathrsfs}
\usepackage{amssymb}
\usepackage{amsmath}
\usepackage{amsthm}
\usepackage{latexsym}
\usepackage{indentfirst}
\usepackage{enumitem}
\usepackage{anysize}
\usepackage{array}
\usepackage{amssymb,amsmath,amsthm,amscd,latexsym}
\usepackage{mathrsfs}
\usepackage{mathrsfs}
\usepackage{amsfonts}
\usepackage{amsmath}
\usepackage{amssymb}
\usepackage{amscd}
\usepackage[all]{xy}
\usepackage{CJK}
\usepackage{xypic}
\usepackage{graphicx}
\usepackage{epstopdf}

\newcommand{\F}{\mathbb{F}}

\newcommand{\Q}{\overline{Q}}
\newcommand{\N}{\overline{N}}

\ifCLASSINFOpdf
\else
\fi
\begin{document}
%
\title{Two new families of two-weight codes$^*$}
%
%
%


\author{Minjia~Shi,
        Yue~Guan,
        and~Patrick~Sol\'e
\thanks{MinJia Shi is with Key Laboratory of Intelligent Computing \& Signal Processing, Ministry of Education, Anhui University No. 3 Feixi Road, Hefei Anhui Province 230039, P. R. China, National Mobile Communications Research Laboratory, Southeast University, 210096, Nanjing, P. R. China and School of Mathematical Sciences of Anhui University, Anhui, 230601, P. R. China (E-mail:smjwcl.good@163.com).}
\thanks{Yue~Guan is with School of Mathematical Sciences of Anhui University, Anhui, 230601, P. R. China.}
\thanks{Patrick~Sol\'e is with CNRS/ LAGA, University Paris 8, 93 526 Saint-Denis, France}
\thanks{Manuscript received November 11, 2016; revised May 14, 2017. M. Shi was supported in part by the National Natural Science Foundation of China
under Grant 61672036, in part by the Technology Foundation
for Selected Overseas Chinese Scholar, Ministry of Personnel, China, under
Grant 05015133, in part by the Open Research Fund of National Mobile
Communications Research Laboratory, Southeast University, under Grant
2015D11, and in part by the Key Projects of Support Program for Outstanding
Young Talents in Colleges and Universities under Grant gxyqZD2016008.}}

\markboth{Journal of \LaTeX\ Class Files,~Vol.~, No.~, ~2017}%
{Shell \MakeLowercase{\textit{et al.}}: Bare Demo of IEEEtran.cls for IEEE Journals}



\maketitle

\begin{abstract}
We construct two new infinite families of trace codes of dimension $2m$, over the ring $\mathbb{F}_p+u\mathbb{F}_p,$ with $u^2=u,$ when $p$ is an odd prime.
They have the algebraic structure of abelian codes. Their Lee weight distribution is computed by using Gauss sums.
 By Gray mapping, we obtain two infinite families of linear $p$-ary codes of respective lengths $(p^m-1)^2$ and $2(p^m-1)^2.$ When $m$ is singly-even, the first family gives five-weight codes.
 When $m$ is odd, and $p\equiv 3 \pmod{4},$ the first family yields
$p$-ary two-weight codes, which are shown to be optimal by application of the Griesmer
bound. The second family consists of two-weight codes that are shown to be optimal, by the Griesmer bound, whenever $p=3$ and $m \ge 3,$ or $p\ge 5$ and $m\ge 4.$ Applications to secret sharing schemes are given.\\
\end{abstract}


%
\IEEEpeerreviewmaketitle

\section{Introduction}

\IEEEPARstart{T}wo-weight codes over fields have been studied since the 1970s due to their connections to strongly regular graphs, finite geometries  and difference sets \cite{D}.
However, most constructions, have used cyclic codes over finite fields \cite{AE,RWM}. In the present paper, we use trace codes over a semi-local ring which is a quadratic extension of
a finite field, and obtain codes over a finite field by Gray mapping. Trace codes are naturally low-rate codes, and not necessarily cyclic. This is part of a general research program where a variety of few weight codes are
obtained by varying the alphabet ring and the defining set \cite{SLS1,SLS2,SLS3}. Here we consider an alphabet ring of odd characteristic, in contrast with \cite{SLS1, SLS2}, and over a non local ring, in contrast with \cite{SLS3}.
We consider two families depending on two different defining sets. These codes are visibly abelian, but possibly not cyclic. The field image of the first family
has two or five weights, depending on the choice of parameters. The second family only contains two-weight codes. Note that abelian codes over rings have been studied
already in \cite{LS}. The alphabet ring we consider here is $\F_p+u\F_p,$ with $u^2=u.$ It is a semi-local ring, which is ring isomorphic to $\F_p\times \F_p$ (Cf. \S 2.1 for a proof).
It is an odd characteristic analogue of
$\F_{2^r}+v\F_{2^r},$ with $v^2=v.$ The latter ring has been employed recently to construct convolutional codes over fields \cite{LSi}.
Some bounds on codes over  $\F_p+u\F_p,$ with $u^2=u$ can be found in \cite{LSi2}.

The defining set of our abelian code is not a cyclic group, but it is an abelian group. The defining set of the first family is related to quadratic
residues in an extension of degree $m$ of $\F_p$, which makes  quadratic Gauss sums appear naturally in the weight distribution analysis, and requires $p$ to be an odd prime.
When $m$ is odd, and $p\equiv 3 \pmod{4},$ we obtain an infinite
family of linear $p$-ary two-weight codes, which are shown to be optimal by application of the Griesmer
bound.  The codes in the second family are also  shown to be optimal by the same technique up to finitely many exceptions.
We show that, both in the five-weight and in the two-weight cases, the first family has a very nice support inclusion structure
which makes it suitable for use in a Massey secret sharing scheme \cite{DY2, KD, YD}. Indeed, we can show that all nonzero codewords are minimal
for the poset of codewords ordered by support inclusion. A similar result holds for the second family. To the best of our knowledge, the weight distributions of the codes obtained here are different from the classical families
of \cite{RWM} and from the those of the codes in \cite{SLS1,SLS2,SLS3}. Our codes are therefore new.

The paper is organized as follows. Section II collects the notions and notations needed in the rest of the article. Section III shows that the trace codes are abelian.
Section IV recalls and reproves some results on Gaussian periods. Section V computes the weight distribution of our codes,
building on the character sum evaluation of the preceding section. Section VI discusses the optimality of the $p$-image from trace codes over $R$. Section VII determines the minimum distance of the dual codes. Section VIII determines the support
structure of the $p$-ary image and describes an application to secret sharing schemes.

\section{Definitions and notations}\subsection{Rings}
Consider the ring $R=\F_p+u\F_p$ where $u^2=u$ and $p$ is a odd prime. It is semi-local with  maximal ideals $(u)$ and $(u-1).$ For any integer $m \ge 1,$ we construct an extension of degree $m$ as ${\mathcal{R}}=\F_{p^m}+u\F_{p^m}$ with again $u^2=u.$
There is a Frobenius operator $F$ which maps $a+ub$ onto $a^p+ub^p.$ The \emph{Trace function,} denoted by $Tr$, is defined as
$Tr=\sum\limits_{j=0}^{m-1}F^j.$
It follows from these definitions that $Tr(a+ub)=tr(a)+utr(b),$ for $a,b \in \F_{p^m}.$
Here $tr()$ denotes the absolute trace of $\F_{p^m},$ given by $$tr(z)=z+z^p+\cdots+z^{p^{m-1}}, z\in\F_{p^m}.$$

The ring ${\mathcal{R} }$ is semi-local with  maximal ideals $(u)$ and $(u-1),$ and respective quotients ${\mathcal{R}}/(u)$ and ${\mathcal{R}}/(u-1)$ are both isomorphic to $\F_{p^m}.$
The Chinese Remainder Theorem shows that $u\F_{p^m}+(1-u)\F_{p^m}$ is isomorphic to the product ring $\F_{p^m}\times\F_{p^m}.$
Similarly, the group of units ${\mathcal{R}}^*$ is  $u\F_{p^m}^*+(1-u)\F_{p^m}^*$ which is isomorphic to $\F_{p^m}^*\times\F_{p^m}^*.$ Here $\F_{p^m}^*$ denotes the multiplicative group of $\F_{p^m}.$
Denote the \emph{squares} and the \emph{non-squares} of $\F_{p^m}$ by ${\mathcal{Q}}$ and ${\mathcal{N}},$ respectively. Thus

$${\mathcal{Q}}=\{ x^2 \mid x \in \F_{p^m}^*\},\, {\mathcal{N}}=\F_{p^m}^*\setminus {\mathcal{Q}}.$$

We write $L=u{\mathcal{Q}}+(1-u)\F_{p^m}^*$ and let $L'={\mathcal{R}}^*$ for simplicity.
Thus $L$ is a subgroup of ${\mathcal{R}}^*$ of index $2.$
\subsection{Gray map}
As a preparation for the image code from trace codes over $R$, we shall take a closer look at the \emph{Gray} map $\phi$ from $R$ to $\F_p^2$, which is defined by $\phi(a+ub)=(-b,2a+b)$ for  $a,b \in \F_p.$ It is a one to one map from $R$ to $\F_p^2,$ which extends naturally into a map from $R^n$ to $\F^{2n}_p.$ Denote the Hamming weight on $\F^{n}_p$ by $w_H(.),$ and the Hamming distance on $\F^{2n}_p$ by $d_H(.,.).$
The \emph{Lee weight} is defined as the Hamming weight of the Gray image $w_L(a+ub)=w_H(-b)+w_H(2a+b)$ for  $a,b \in \F_p^n.$
The \emph{Lee distance} of $x,y\in R^n$ is defined as $d_L(x,y)=w_L(x-y).$
Thus the Gray map is a linear isometry from $(R^n,d_L)$ to $({\F^{2n}_p},d_H).$
For convenience, we write $N=2n$ in the rest of the paper.
\subsection{Codes}
 A {\bf linear code} $C$ over $R$ of length $n$ is an $R$-submodule of $R^n$. If $x=(x_1,x_2,\cdots,x_n)$
 and $y=(y_1,y_2,\cdots,y_n)$ are two elements of  $R^n$, their standard inner product
  is defined by $\langle x,y\rangle=\sum\limits_{i=1}^nx_iy_i$, where the operation is performed in $R$. The {\bf dual code} of $C$ is denoted by $C^\perp$ and
  defined as $C^\perp=\{y\in R^n|\langle x,y\rangle =0, \forall x\in C\}.$ By definition, $C^\perp$ is also a linear code over $R$.
  Given a finite abelian group $G,$ a code over $R$ is said to be {\bf abelian} \cite{B}, if it is an ideal of the group ring $R[G].$
  Recall that the ring $R[G]$ is defined on functions from $G$ to $R$ with pointwise addition as addition, and convolution product as multiplication.
  Concretely, it is the set of all formal sums $f=\sum_{h \in G} f_h X^h,$ with addition and multiplication defined as follows.
  If $f,g\in R[G],$  we write
  $$f+g=\sum_{g \in G} (f_h+g_h) X^h,$$ and

  $$ fg =\sum_{h \in G} (\sum_{r+s=h} f_r g_s)X^h.$$
  In other words, the coordinates of $C$ are indexed by elements of $G$ and $G$ acts regularly on this set. For more details on abelian codes see \cite{W}.
  In the special case when $G$ is cyclic, the code is a cyclic code in the usual sense \cite{MS}.

\section{Symmetry}
For $a\in \mathcal{R}$, define the vector $ev(a)$ by the following evaluation map
$ev(a)=(Tr(ax))_{x\in L}.$ Define the code $C(m,p)$ by the formula $C(m,p)=\{ev(a)|a\in \mathcal{R}\}$. Thus $C(m,p)$ is a code of length $|L|=\frac{(p^{m}-1)^2}{2}$ and size $\vert R\vert ^m$ over $R.$
Similarly, define the vector $ev'(a)$ by the following evaluation map
$ev'(a)=(Tr(ax))_{x\in L'},$ and  the code $C'(m,p)$ by the formula $C'(m,p)=\{ev'(a)|a\in \mathcal{R}\}.$
Thus $C'(m,p)$ is a code of length $|L'|=(p^{m}-1)^2$ and size $\vert R\vert ^m$ over $R.$

\textbf{Proposition 3.1} The group $L$ $($resp. $L'$ $)$ acts regularly on the coordinates of $C(m,p)$ $($resp. $C'(m,p)$ $)$.

\textbf{Proof} For any $w,v \in L$ the change of variables $ x\mapsto (v/w)x$ maps $w$ to $v.$ This transformation defines thus a transitive action of $L$ on itself. Given an ordered pair $(w,v)$ this transformation is unique,
hence the action is regular. A similar argument holds for $C'(m,p)$ and $L'.$


The code $C(m,p)$ is thus an {\em abelian code} with respect to the group $\mathcal{R}^*.$
In other words, it is an ideal of the group ring $R[\mathcal{R}^*].$ As observed in the previous section $\mathcal{R}^*$ is not a cyclic group, hence $C(m,p)$ may be not cyclic.
\section{Character sums}
In this section we give some background material on character sums.
Let $\chi$ denote an arbitrary multiplicative character of $\F_q.$ Assume $q$ is odd.
Denoted by $\eta$ the quadratic multiplicative character is defined by $\eta(x)=1,$ if $x$ is a square and $\eta(x)=-1,$ if not.
Let $\psi$ denote the standard canonical additive character of $\F_q.$ The \emph{squares} and the \emph{non-squares} of $\F_{q}$ are denoted, extending the notation of \S 2.1,
by ${\mathcal{Q}}$ and ${\mathcal{N}},$ respectively. Thus,

$${\mathcal{Q}}=\{ x^2 \mid x \in \F_{q}^*\},\, {\mathcal{N}}=\F_{q}^*\setminus {\mathcal{Q}}.$$
The classical {\bf Gauss sum} can be defined as $G(\chi)=\sum\limits_{x\in \F_q^*}\psi(x)\chi(x).$

We define the following character sums
\begin{eqnarray*}
\Q=\sum_{x\in {\mathcal{Q}}}\psi(x), \ \N=\sum_{x\in {\mathcal{N}}}\psi(x).
\end{eqnarray*}
On the basis of orthogonality of characters \cite[Lemma 9, p. 143]{MS} it is evident that $\Q+\N=-1.$
Noting that the characteristic function of ${\mathcal{Q}}$ is $\frac{1+\eta}{2},$ we get then
\begin{eqnarray*}
\Q=\frac{G(\eta)-1}{2}, \ \N=\frac{-G(\eta)-1}{2}.
\end{eqnarray*}

It is well known \cite{DY} that if $q=p^m,$ the quadratic Gauss sums can be evaluated as
\begin{eqnarray}
G(\eta)&=&(-1)^{m-1}\sqrt{q}\,,\ p\equiv 1 \pmod{4},\\
G(\eta)&=&(-1)^{m-1}i^m\sqrt{q}, \,p\equiv 3 \pmod{4}. \label{im}
\end{eqnarray}

Particularly, if $m$ is singly-even, these formulas can be simplified to
$G(\eta)=\epsilon(p)\sqrt{q},$ with $\epsilon(p)=(-1)^{\frac{(p+1)}{2}},$
yielding
\begin{eqnarray*}
\Q=\frac{\epsilon(p)\sqrt{q}-1}{2}, \N=-\frac{\epsilon(p)\sqrt{q}+1}{2}.
\end{eqnarray*}
In fact $\Q$ and $\N$ are examples of \emph{Gaussian periods}, and these relations could have been deduced from \cite[Lemma 11]{DY}.

\section{Weight distributions of trace codes}

Let $\omega=\exp(\frac{2\pi i}{p})$ be a complex root of unity of order $p.$ If $y=(y_1,y_2,\cdots,y_N)\in \mathbb{F}_p^N,$ let $\Theta(y)=\sum\limits_{j=1}^N\omega^{y_j}.$
For simplicity, we let $\theta(a)=\Theta(\phi(ev(a))).$ By linearity of the Gray map, and of the evaluation map, we see that $\theta(sa)=\Theta(\phi(ev(sa))),$ for any $s\in \F_p^*.$ For our purpose, let us begin with the following correlation lemma.

\textbf{Lemma 5.1}\label{3.1} \cite{SLS2} For all $y=(y_1,y_2,\cdots,y_N)\in \mathbb{F}_p^N,$ we have
$\sum\limits_{s=1}^{p-1}\Theta(sy)=(p-1)N-pw_H(y).$

In connection with the proceding discussion, we now distinguish two cases of weight distributions depending on the defining set.
\subsection{The case of $L={\mathcal{Q}}\times \F_{p^m}^*$}
\subsubsection{$m$ is singly-even}

\textbf{Theorem 5.2}\label{enum1} Assume $m$ is singly-even. For $a\in \mathcal{R}$, the Lee weight of codewords of $C(m,p)$ is given below.

\begin{enumerate}
\item[(a)] If $a=0$, then $w_L(ev(a))=0$;
\item[(b)] If $a=u\alpha$, $\alpha\in\mathbb{F}_{p^m}^*,$ then if \begin{enumerate}
\item[$\alpha \in {\mathcal{Q}}$] then $w_L(ev(a))=(p-1)\big(p^{2m-1}-p^{m-1}-\epsilon(p)p^{{3m}/2-1}+\epsilon(p)p^{m/2-1}\big)$,
\item[$\alpha \in {\mathcal{N}}$] then $w_L(ev(a))=(p-1)\big(p^{2m-1}-p^{m-1}+\epsilon(p)p^{{3m}/2-1}-\epsilon(p)p^{m/2-1}\big);$
\end{enumerate}
\item[(c)] If $a=(1-u)\beta$, $\beta\in\mathbb{F}_{p^m}^*,$ then $w_L(ev(a))=(p-1)(p^{2m-1}-p^{m-1})$;
\item[(d)] If $a=u\alpha+(1-u)\beta\in \mathcal{R}^*$, then if \begin{enumerate}
\item[$\alpha \in {\mathcal{Q}}$] then $w_L(ev(a))=(p-1)\big(p^{2m-1}-2p^{m-1}+\epsilon(p)p^{m/2-1}\big)$,
\item[$\alpha \in {\mathcal{N}}$] then  $w_L(ev(a))=(p-1)\big(p^{2m-1}-2p^{m-1}-\epsilon(p)p^{m/2-1}\big).$
\end{enumerate}
 \end{enumerate}

\textbf{Proof}
\begin{enumerate}
\item[(a)] If $a=0$, then $Tr(ax)=0$. So $w_L(ev(a))=0$.
\item[(b)] If $a=u\alpha,\,x=ut+(1-u)t'$ with $\alpha \in \F_{p^m}^*,$ then $ax= u\alpha t,$ and $Tr(ax)=Tr(u\alpha t)=utr(\alpha t).$
Taking Gray map yields
$\phi(ev(a))=(-tr(\alpha t),tr(\alpha t))_{t,t'}.$
Taking character sums
\begin{eqnarray*}
\theta(a)&=&\sum_{t\in {\mathcal{Q}}}\sum_{t'\in \F_{p^m}^*}\omega^{-tr(\alpha t)}+\sum_{t\in {\mathcal{Q}}}\sum_{t'\in \F_{p^m}^*}\omega^{tr(\alpha t)}\\ &=&2\sum_{t\in {\mathcal{Q}}}\sum_{t'\in \F_{p^m}^*}\omega^{-tr(\alpha t)}\\
&=&2(p^m-1)\sum_{t\in {\mathcal{Q}}}\omega^{tr(\alpha t)}.
\end{eqnarray*}
Replaced $\alpha t$ by $t$, it is easy to check that the last character sum is $\Q$ or $\N$ depending on $\alpha \in {\mathcal{Q}}$ or $\alpha \in {\mathcal{N}}.$
Since $m$ is even, $s\in \F_p^*$ is a square in $\F_{p^m}.$ Thus $\theta(sa)=\theta(a),$ for any $s\in \F_p^*.$ The statement follows from Lemma 5.1.
Thus $w_L(ev(a))=\frac{p-1}{p}(N-2(p^m-1)\Q),$ or $w_L(ev(a))=\frac{p-1}{p}(N-2(p^m-1)\N),$ according to the value of $\eta(\alpha).$
\item[(c)] If $a=(1-u)\beta,\,x=ut+(1-u)t'$ with $\beta \in \F_{p^m}^*,$ then $ax=\beta t'-u\beta t',$ and $Tr(ax)=tr(\beta t')-utr(\beta t').$
Taking Gray map yields
$\phi(ev(a))=(tr(\beta t'),tr(\beta t'))_{t,t'}.$
Taking character sums
$\theta(a)=2\sum\limits_{t\in {\mathcal{Q}}}\sum\limits_{t'\in \F_{p^m}^*}\omega^{tr(\beta t')}=1-p^m.$ Thus $w_L(ev(a))=(p-1)(p^{2m-1}-p^{m-1}).$
\item[(d)] Let $a=u\alpha+(1-u)\beta\in R^*,\,x=ut+(1-u)t'.$ So $Tr(ax)=tr(\beta t')+utr(\alpha t-\beta t').$ Thus
$\phi(ev(a))=(-tr(\alpha t-\beta t'),tr(\alpha t+\beta t'))_{t,t'}$ by the Gray map.
Taking character sums
$\theta(a)=\sum\limits_{t\in \mathcal{Q}}\omega^{-tr(\alpha t)} \sum\limits_{t'\in \F_{p^m}^*}\omega^{tr(\beta t')}+\sum\limits_{t\in \mathcal{Q}}\omega^{tr(\alpha t)} \sum\limits_{t'\in \F_{p^m}^*}\omega^{tr(\beta t')}=-2\sum\limits_{t\in \mathcal{Q}}\omega^{tr(\alpha t)}.$
By a change of variable $t=\alpha t,$ we see that the last character sum is $\Q$ or $\N$ depending on $\alpha \in {\mathcal{Q}}$ or $\alpha \in {\mathcal{N}}.$
 Thus $w_L(ev(a))=\frac{p-1}{p}(N+2\Q),$ or $w_L(ev(a))=\frac{p-1}{p}(N+2\N),$ considering the value of $\eta(\alpha).$
\end{enumerate}

Therefore, we have constructed a $p$-ary code of length $N=(p^m-1)^2,$ dimension $2m,$ with five weights. The weight distribution is given in Table I.
\begin{center}$\mathrm{Table~ I. }~~~\mathrm{weight~ distribution~ of}~ C(m,p) $ in Theorem 5.2\vspace{0.2cm}

\begin{tabular}{cc||cc}
\hline
  Weight && Frequency  \\
  \hline

  0         && 1\\
  $(p-1)(p^{m-1}-p^{m/2-1})(p^m-1)$          &&              $\frac{p^m-1}{2}$\\
  $(p-1)(p^{2m-1}-2p^{m-1}-p^{{m/2}-1})$    && $\frac{(p^m-1)^2}{2}$\\
  $(p-1)(p^{2m-1}-2p^{m-1}+p^{{m/2}-1})$  &&   $\frac{(p^m-1)^2}{2}$ \\
  $(p-1)(p^{2m-1}-p^{m-1})$   && $p^m-1$ \\
  $(p-1)(p^{m-1}+p^{m/2-1})(p^m-1)$    &&  $\frac{p^m-1}{2}$\\
  \hline
\end{tabular}
\end{center}
(Note that taking $\epsilon(p)=1,$ or $-1,$ leads to the same values.)

\subsubsection{$m$ is odd and $p\equiv 3 \pmod{4}$}
Note that in that case by (\ref{im}) in Section IV we see that $G(\eta)$ is imaginary. This implies that $\Re(\Q)=\Re(\N)=-\frac{1}{2},$ where $\Re(z)$ denotes the real part of the complex number $z.$ We need first to refine the following correlation lemma.

\textbf{Lemma 5.3} \label{trick} {\cite{SLS2} If $p\equiv 3 \pmod{4},$ then we have $\sum\limits_{s=1}^{p-1}\theta(sa)=(p-1)\Re(\theta(a)).$}

\textbf{Theorem 5.4}\label{enum2} Assume $m$ is odd and $p\equiv 3 \pmod{4}.$ For $a\in \mathcal{R}$, the Lee weight of codewords of $C(m,p)$ is given below.
\begin{enumerate}
\item[(a)] If $a=0$, then $w_L(ev(a))=0$;
\item[(b)] If $a=u\beta$ with $\beta\in \F_{p^m}^*,$ then $w_L(ev(a))=(p-1)(p^{2m-1}-p^{m-1})$;
\item[(c)] If $a=(1-u)\beta$ with $\beta\in \F_{p^m}^*,$ then $w_L(ev(a))=(p-1)(p^{2m-1}-p^{m-1})$;
\item[(d)] If $a\in \mathcal{R}^*$, then $w_L(ev(a))=(p-1)(p^{2m-1}-2p^{m-1})$.
 \end{enumerate}

\textbf{Proof}
 The proof of the case (a) is like that of Theorem 5.2. The case (b) follows from Lemma 5.3 applied to the correlation lemma.
Thus  $\Re(\theta(a))=1-p^m,$ and $pw_L(ev(a))=(p-1)(N-\Re(\theta(a))),$ yielding $w_L(ev(a))=(p-1)(p^{2m-1}-p^{m-1}).$
The result follows.
The proof of case (c) is the same as that of case (b).
In the case (d), $\Re(\theta(a))=1$, then $w_L(ev(a))=(p-1)(p^{2m-1}-2p^{m-1}).$

Thus we obtain a family of $p$-ary two-weight codes of parameters $[p^{2m}-2p^{m}+1,2m],$ with weight distribution as given in Table II. The parameters are different from
those in \cite{RWM}, \cite{SLS1}, \cite{SLS2} and \cite{SLS3}.
\begin{center}$\mathrm{Table~ II. }~~~\mathrm{weight~ distribution~ of}~ C(m,p)$ in Theorem 5.4\vspace{0.2cm}
\begin{tabular}{ccc||cc}
\hline
  Weight  \  \  & &&& \  \ Frequency  \\
  \hline

  0  \  \      &&&  & \  \ 1\\
  $(p-1)(p^{2m-1}-2p^{m-1})\  \ $                       &&&  & \  \ $(p^{m}-1)^2$\\
  $(p-1)(p^{2m-1}-p^{m-1})\  \ $                     &&&  & \  \ $2(p^m-1)$\\
  \hline
\end{tabular}
\end{center}
\subsection{The case of $L'={\F_{p^m}^* \times \F_{p^m}^*}$}

\textbf{Theorem 5.5}\label{enum} For $a\in \mathcal{R}$, the Lee weight of codewords of $C'(m,p)$ is
\begin{enumerate}
\item[(a)] If $a=0$, then $w_L(ev'(a))=0;$
\item[(b)] If $a=u\alpha$, $\alpha\in\mathbb{F}_{p^m}^*,$ then $w_L(ev'(a))=2(p-1)(p^{2m-1}-p^{m-1});$
\item[(c)] If $a=(1-u)\beta$, $\beta\in\mathbb{F}_{p^m}^*,$ then $w_L(ev'(a))=2(p-1)(p^{2m-1}-p^{m-1});$
\item[(d)] If $a\in \mathcal{R}^*$, then $w_L(ev'(a))=2(p-1)(p^{2m-1}-2p^{m-1}).$
 \end{enumerate}

\textbf{Proof}
\begin{enumerate}
\item[(a)] If $a=0$, then $Tr(ax)=0$. So $w_L(ev'(a))=0$.
\item[(b)] If $a=u\alpha,\,x=ut+(1-u)t'$ with $\alpha \in \F_{p^m}^*,$ then $ax= u\alpha t,$ and $Tr(ax)=Tr(u\alpha t)=utr(\alpha t).$
Taking Gray map yields
$\phi(ev'(a))=(-tr(\alpha t),tr(\alpha t))_{t,t'}.$
Taking character sums
$\theta(a)=\sum\limits_{t \in \F_{p^m}^*}\sum\limits_{t'\in \F_{p^m}^*}\omega^{-tr(\alpha t)}+\sum\limits_{t\in \F_{p^m}^*}\sum\limits_{t'\in \F_{p^m}^*}\omega^{tr(\alpha t)}=2\sum\limits_{t\in \F_{p^m}^*}\sum\limits_{t'\in \F_{p^m}^*}\omega^{-tr(\alpha t)}=-2(p^m-1).$
Thus $w_L(ev'(a))=2(p-1)(p^{2m-1}-p^{m-1}).$
\item[(c)] If $a=(1-u)\beta,\,x=ut+(1-u)t'$ with $\beta \in \F_{p^m}^*,$ then $ax=\beta t'-u\beta t',$ and $Tr(ax)=tr(\beta t')-utr(\beta t').$
Taking Gray map yields
$\phi(ev'(a))=(tr(\beta t'),tr(\beta t'))_{t,t'}.$
Taking character sums
$\theta(a)=2\sum\limits_{t\in  \F_{p^m}^*}\sum\limits_{t'\in \F_{p^m}^*}\omega^{tr(\beta t')}=2-2p^m.$ Thus $w_L(ev'(a))=2(p-1)(p^{2m-1}-p^{m-1}).$
\item[(d)] Let $a=u\alpha+(1-u)\beta),\,x=ut+(1-u)t'.$ So $ax=\beta t'+u(\alpha t-\beta t'),$ and $Tr(ax)=tr(\beta t')+utr(\alpha t-\beta t').$ Taking Gray map yields
$\phi(ev'(a))=\big(-tr(\alpha t-\beta t'),tr(\alpha t+\beta t')\big)_{t,t'}.$
Taking character sums
$\theta(a)=\sum\limits_{t\in  \F_{p^m}^*}\omega^{-tr(\alpha t)} \sum\limits_{t'\in \F_{p^m}^*}\omega^{tr(\beta t')}+\sum\limits_{t\in \F_{p^m}^*}\omega^{tr(\alpha t)} \sum\limits_{t'\in \F_{p^m}^*}\omega^{tr(\beta t')}=2.$
 Thus $w_L(ev'(a))=\frac{p-1}{p}(N'-2),$ or $w_L(ev'(a))=2(p-1)(p^{2m-1}-2p^{m-1}).$
\end{enumerate}

Thus we have constructed a $p$-ary code of length $N'=2(p^m-1)^2,$ dimension $2m,$ with two weights. The weight distribution is given in Table III. Note that the parameters are different from those in
[4], [14], [15] and [16].
\begin{center}$\mathrm{Table~ III. }~~~\mathrm{weight~ distribution~ of}~ C'(m,p) $ in Theorem 5.5\vspace{0.2cm}
\begin{tabular}{ccc||c}
\hline
  Weight  && &   Frequency  \\
  \hline

  0     & &  &    1 \\
  $2(p-1)(p^{2m-1}-2p^{m-1})$         &         &        &  $p^{2m}-2p^{m}+1$\\
  $2(p-1)(p^{2m-1}-p^{m-1})$    &            &         & $2p^m-2$\\
  \hline
\end{tabular}
\end{center}

\section{Optimality of the $p$-ary image}
A central question of coding theory is to decide whether the constructed codes are optimal or not. In this section, we will investigate the optimality of the $p$-ary image of the trace codes we constructed in Section V.
Recall the $p$-ary version of the Griesmer bound. If $[N,K,d]$ are the parameters of a linear $p$-ary code, then
$$\sum_{j=0}^{K-1}\bigg\lceil \frac{d}{p^j} \bigg\rceil \le N.$$
\subsection{$L=u{\mathcal{Q}}+(1-u)\F_{p^m}^*,$ $m$ is odd and $p\equiv 3 \pmod{4}$}

\textbf{Theorem 6.1}
Assume $m$ is odd and $m\geq 3$, and $p\equiv 3 \pmod{4}.$ The code $\phi(C(m,p))$ is optimal.

\textbf{Proof}
Firstly, $N=p^{2m}-2p^{m}+1,\, K=2m,\,d=(p-1)(p^{2m-1}-2p^{m-1})$ on account of Theorem 5.4. We claim that $\sum\limits_{j=0}^{K-1}\big\lceil \frac{d+1}{p^j}\big\rceil >N,$ contradicting the Griesmer bound. The ceiling function takes three values depending on $j$.
\begin{itemize}
 \item $0\leq j\leq m-1 \Rightarrow \lceil \frac{d+1}{p^j} \rceil =p^{2m-j}-2p^{m-j}-p^{2m-j-1}+2p^{m-j-1}+1$;
 \item $j= m \Rightarrow \lceil \frac{d+1}{p^j} \rceil =p^m-p^{m-1}-1$;
 \item $m<j\leq2m-1 \Rightarrow \lceil \frac{d+1}{p^j} \rceil =p^{2m-j}-p^{2m-j-1}$.
\end{itemize}
Thus $\sum\limits_{j=0}^{K-1}\big\lceil \frac{d+1}{p^j} \big\rceil=p^{2m}-2p^m+m.$
Note that $\sum\limits_{j=0}^{K-1}\lceil \frac{d+1}{p^j} \rceil-N=m-1>0.$

\textbf{Example 6.2}
Let $p=3$ and $m=3$, we obtain a ternary code of parameters $[676, 6, 450].$ The weights of this code are 450 and 468 with frequencies 676 and 52, respectively.

\subsection{$L'={\mathcal{R}}^*,$ $m$ is an arbitrary integer and $p$ is odd prime}

\textbf{Theorem 6.3}
Assume $m\geq 3$ and $p=3$ or $m\geq 4$ and the odd prime $p\geq 5$. The code $\phi(C'(m,p))$ is optimal.

\textbf{Proof}
It follows from Theorem 5.5 that $N'=2(p^{2m}-2p^{m}+1),\, K=2m,\,d=2(p-1)(p^{2m-1}-2p^{m-1}).$ We claim that $\sum\limits_{j=0}^{K-1}\big\lceil \frac{d+1}{p^j} \big\rceil >N',$ violating the Griesmer bound. The ceiling function takes the following values depending on the position of $j$.
\begin{enumerate}
\item[(a)] when $p=3$, the ceiling function takes three
values depending on $j$.
\begin{itemize}
 \item $0\leq j\le m-1 \Rightarrow \lceil \frac{d+1}{p^j} \rceil =2p^{2m-j}-4p^{m-j}-2p^{2m-j-1}+4p^{m-j-1}+1$;
 \item $j= m \Rightarrow \lceil \frac{d+1}{p^j} \rceil =2p^m-2p^{m-1}-2$;
 \item $m<j\leq 2m-1 \Rightarrow \lceil \frac{d+1}{p^j} \rceil =2p^{2m-j}-2p^{2m-j-1}$.
\end{itemize}
Thus $\sum\limits_{j=0}^{K-1} \big\lceil \frac{d+1}{p^j} \big\rceil=2p^{2m}-4p^m+m.$
Note that $\sum\limits_{j=0}^{K-1}\big\lceil \frac{d+1}{p^j} \big\rceil-N'=m-2>0.$
\item[(b)] when $p\geq 5$ and $p$ is odd prime, the ceiling function takes three
values depending on $j$.
\begin{itemize}
 \item $0\leq j\le m-1 \Rightarrow \lceil \frac{d+1}{p^j} \rceil =2p^{2m-j}-4p^{m-j}-2p^{2m-j-1}+4p^{m-j-1}+1$;
 \item $j= m \Rightarrow \lceil \frac{d+1}{p^j} \rceil =2p^m-2p^{m-1}-3$;
 \item $m<j\leq 2m-1 \Rightarrow \lceil \frac{d+1}{p^j} \rceil =2p^{2m-j}-2p^{2m-j-1}$.
\end{itemize}
Thus $\sum\limits_{j=0}^{K-1}\big\lceil \frac{d+1}{p^j} \big\rceil=2p^{2m}-4p^m+m-1.$
Note that $\sum\limits_{j=0}^{K-1}\big\lceil \frac{d+1}{p^j} \big\rceil-N'=m-3>0.$
\end{enumerate}
Hence the theorem is proved.

\textbf{Example 6.4}
Let $p=3$ and $m=3$, we obtain a ternary code of parameters $[1352, 6, 900].$ The weights of this code are 900 and 936 with frequencies 676 and 52, respectively.

\section{The minimum distance of the dual code}
We compute the dual distance of $\phi(C(m,p))({\rm resp}. \ \phi(C'(m,p))).$ In connection with the discussion in \cite{SLS1}, we mention without proof the following lemma.

\textbf{Lemma 7.1}\label{nonde}
 If for all $a \in \mathcal{R},$ we have that $Tr(ax)=0,$ then $x=0.$

\textbf{Theorem 7.2} \label{d'}
 For all odd primes $p$ and all $m\ge 2,$ the dual Lee distance $d'$ of $C(m,p)$ is $2.$

\textbf{Proof}
 First, we check that $d'\ge 2$ by showing that $C(m,p)^\bot$ does not contain a word of Lee weight one.
 If it does, let us assume first that it has value $\gamma \neq 0$ at some $x \in L.$ This implies that $\forall a \in \mathcal{R},\gamma
 Tr(ax)=0$ or $Tr(a \gamma x)=0,$ and by Lemma 7.1 $\gamma x=0.$ Contradiction with $\gamma \neq 0.$
 If that word takes the value $\gamma (1-2u)$ at some $x \in L,$ then writing $x=ut+(1-u)t'$ and $a=u\alpha+(1-u)\beta,$ with $\alpha,\beta$ in $\F_{p^m},$ with $t\in \mathcal{Q},t'\in \F_{p^m}^*$ yields,
 after reduction modulo $M$ the equation  valid $\forall \alpha, \beta \in \F_{p^m},$
 $\gamma tr(\beta t')=0$ and $tr(\gamma\alpha t)=0,$ and we conclude, by the nondegenerate character of $tr(),$ that $\gamma t=\gamma t'=0.$ Contradiction with $x\neq 0$.

 Next, we shall show that $d'<3.$ If not, we can apply the sphere-packing bound to $\phi(C(m,p))^\bot,$ to obtain
 $p^{2m}\ge 1+N(p-1)=1+(p^{2m}-2p^{m}+1)(p-1),$ or, after expansion
 $2p^{2m}+2p^{m+1}\ge p+p^{m}(p^{m+1}+2).  $
 Dropping the $p$ in the RHS, and dividing both sides by $p^m,$ we find that this inequality would imply $p^{m}< f(p),$ with $f(x)=\frac{2x-2}{x-2}=2+\frac{2}{x-2}.$
 But the function $f$ is decreasing for $x\geq3,$ and $f(3)=4,$ while $p^m\ge p^2\ge 9.$ Contradiction.

We now consider codes from the second family.

\textbf{Theorem 7.3} \label{d}
 For all odd prime $p$ and all $m\ge 2,$ the dual Lee distance $d'$ of $C'(m,p)$ is $2.$

\textbf{Proof}
 First, we check that $d'\ge 2.$ The proof of $d'\ge 2$ is like that in Theorem 7.2.
 Next, we show that $d'<3.$ Otherwise, we can apply the sphere-packing bound to $\phi (C'(m,p)^\bot),$ to obtain
 $$p^{2m}\ge 1+N'(p-1)=1+2(p^{2m}-2p^{m}+1)(p-1).$$ It is shown by a straightforward argument that $1+N'(p-1)>1+N'$ where $p$ is an odd prime. Then we check $p^{2m}>1+N'.$ We obtain the inequality $(p^m-3)(p^m-1)<0$ after calculation which is contradict $p^m\geq3.$ Hence $p^{2m}\leq 1+N'(p-1)=1+2(p^{2m}-2p^{m}+1)(p-1).$

\section{Applications to secret sharing schemes}
To illustrate an application of our constructed codes in secret sharing, we review the fundamentals of this cryptographic protocol.
The concept of secret sharing schemes was first proposed by Blakley and Shamir in 1979. We present some basic definitions concerning secret sharing schemes and refer the interested reader to the survey \cite{DR} for details.

The sets of participants which are capable of recovering the secret $S$ are called \emph{access sets}. The set of all access sets is called the \emph{access structure} of the scheme. An access set is called \emph{minimal} if its members can recover the secret $S$ but the members of any of its proper subsets cannot recover $S$. Furthermore, if a participant is contained in every minimal access set in the scheme, then it is a  dictatorial participant.

The support $s(x)$ of a vector $x$ in $\F_p^N$ is defined as the set of indices where it is nonzero. We say that a vector $x$ covers a vector $y$ if $s(x)$ contains $s(y).$
A \emph{minimal codeword} of a linear code $C$ is a nonzero codeword that does not cover any other nonzero codeword.

Next, we recall the secret sharing scheme based on linear codes which is constructed by Massey. Let $C$ be an $[n,k]$ linear code over the given finite field $\F_p$ and $G=[g_0,g_1,\cdots,g_{n-1}]$ be a generator matrix of $C$ where the column vectors are nonzero. The dealer chooses a random vector $u=(u_0,u_1,\cdots,u_{k-1})\in\F_p^k$ and encodes the chosen vector as $c=uG=(c_0,c_1,\cdots,c_{n-1}).$ Then the dealer keeps the value of $u$ at the first coordinate $S=c_0=ug_0$ as a \emph{secret}, and distributes the values at the remaining coordinates of $c$ to the participants as \emph{shares}. Note that $S=c_0=ug_0,$ the set of shares $\{v_{i_1}=c_{i_1},\cdots,v_{i_t}=c_{i_t}\}$ can recover the secret $S$ if and only if the vectors $g_0, g_{i_1},\cdots,g_{i_t}$ are linearly independent. Hence, we can write the secret as a linear combination $$S=ug_0=\sum_{j=1}^tx_jc_{i_j}.$$ From this equation, it is clear that if we have shares $c_{i_j}, 1\leq j\leq t$ and find $x_j$ by $g_0=\sum_{j=1}^tx_jg_{i_j}$, we can recover the secret $S.$ In fact, we can use the participants that correspond to the nonzero coordinates of the minimal codewords $v\in C^\bot$, because $cv=c_0+c_{i_1}v_{i_1}+\cdots+c_{i_t}v_{i_t}$ which implies that $S=c_0=-(c_{i_1}v_{i_1}+\cdots+c_{i_t}v_{i_t})$.

In general, it is a tough task to determine the minimal codewords of a given linear code. However, there is a numerical condition, derived in \cite{AB}, bearing on the weights of the code, that is easy to check.

\textbf{Lemma 8.1} (Ashikmin-Barg) Let $w_0$ and $w_{\infty}$ denote the minimum and maximum nonzero weights, respectively. If
$\frac{w_0}{w_{\infty}}>\frac{p-1}{p},$ then every nonzero codeword of $C$ is minimal.

In the special case when all nonzero codewords are minimal, it was shown in \cite{DY2} that there is the following alternative, depending on $d'$.

\textbf{Lemma 8.2} (\cite{DY2})\label{abcd}
Let $C$ be an $[n,k]$ code over $\F_p$ and $G=[g_0,g_1,\cdots,g_{n-1}]$ be a generator matrix of $C$. If every codeword of $C$ is minimal vector, then there are $p^{k-1}$ minimal access sets and the total number of participants is $n-1$ in the secret sharing scheme based on $C^\bot.$ Let $d'$ denote the minimal distance of $C^\bot$. We have the following results:
\begin{itemize}
 \item If $d'\ge 3,$ then for any fixed $1\leq t'\leq min\{k-1,d'-2\},$ every group of $t'$ participants is involved in $(p-1)^{t'}p^{k-(t'+1)}$ out of $p^{k-1}$ minimal access sets. We call such participants dictators.
 \item When $d'=2$, if $g_i$ is a multiple of $g_0$, $1\leq i\leq n-1$, then the participant $P_i$ must be in every minimal access set and such a participant is called a dictatorial participant. If $g_i$ is not a multiple of $g_0$, then the participant $P_i$ must be in $(p-1)p^{k-2}$ out of $p^{k-1}$ minimal access sets.
\end{itemize}

\textbf{Theorem 8.3}
Let $G=[g_0,g_1,...,g_{(p^m-1)^2-1}]$ be a generator matrix of $\phi (C(m,p))$ where $m(\geq 3)$ is odd and $p\equiv 3 \pmod{4}$, then there are $p^{2m-1}$ minimal access sets and the total number of participants is $(p^m-1)^2-1$ in the secret sharing scheme based on $\phi (C(m,p)^\bot).$ We have the following results:
\begin{itemize}
 \item if $g_i$ is a multiple of $g_0$, $1\leq i\leq (p^m-1)^2-1$, then the participant $P_i$ must be in every minimal access set and such a participant is called a dictatorial participant.
 \item If $g_i$ is not a multiple of $g_0$, then the participant $P_i$ must be in $(p-1)p^{2m-2}$ out of $p^{2m-1}$ minimal access sets.
\end{itemize}

\textbf{Proof}
By the preceding lemma with $w_0=(p-1)(p^{2m-1}-2p^{m-1})$ and $w_{\infty}=(p-1)(p^{2m-1}-p^{m-1})$ in Table II. Rewriting the inequality of the lemma as $pw_0>(p-1)w_{\infty},$ and dividing both sides by $\frac {p^m}{p-1},$ we obtain $ p(p^{m}-2)>(p-1)(p^{m}-1),$ or $p^m-p-1>0,$ which is true for $m\ge 3.$
Substitute $n=(p^m-1)^2$ into Lemma 8.2, and then the conclusion is obtained.

\textbf{Remark 8.4}
By the similar method in the proof of Theorem 8.3, we find that all the nonzero codewords of $\phi(C(m,p))$ ($m(\geq 6)$ is singly-even) and $\phi(C'(m,p))$ ($m \ge 2, p$ is odd prime) are minimal. Thus, when $m(\geq 6)$ is singly-even, the secret sharing scheme based on $\phi(C(m,p)^\bot)$ has a similar structure as that in the case when $m\geq 3$ is odd and $p\equiv 3 \pmod{4}.$ Let $G'=[g_0,g_1,...,g_{2(p^m-1)^2-1}]$ be a generator matrix of $\phi (C'(m,p))$, then there are $p^{2m-1}$ minimal access sets and the total number of participants is $2(p^m-1)^2-1$ in the secret sharing scheme based on $\phi (C'(m,p)^\bot).$ If $g_j$ is a multiple of $g_0$, $1\leq j\leq 2(p^m-1)^2-1$, then the participant $P_j$ must be in every minimal access set and such a participant is called a dictatorial participant. If $g_j$ is not a multiple of $g_0$, then the participant $P_j$ must be in $(p-1)p^{2m-2}$ out of $p^{2m-1}$ minimal access sets.

\section{Conclusion}
In the present paper, we have studied two infinite families of trace codes defined over a finite ring. Because the defining sets of these codes have the structure of abelian
multiplicative groups, they inherit the structure of abelian codes. It is an open problem to determine if they are cyclic codes or not. More importantly,
it is worthwhile to study other defining sets that are subgroups of the group of units
of $\mathcal{R}.$ In particular, it would be interesting to replace our Gaussian periods $\Q,\N$  by other character sums that are amenable to exact evaluation, in the vein of the sums which appear in the study of irreducible
cyclic codes \cite{DY,MR}. This would lead to other enumerative results of codes with a few weights.

Compared to the codes we constructed by similar techniques in \cite{SLS1}, \cite{SLS2} and \cite{SLS3}, the obtained codes in
this paper have different weight distributions.  They are also different from the weight distributions in the classical families in \cite{RWM}. Hence,
the $p$-ary linear codes constructed from trace codes over rings in this paper are new.

\textbf{Biographies of all authors}

Minjia Shi, he received the B.S. degree in Mathematics from Anqing Normal College of China in 2004; the M.S. degree in Mathematics from Hefei University of Technology of China in 2007, and the Ph.D degree in the Institute of Computer Network Systems from Hefei University of Technology of China in 2010. He has been teaching at the School of Mathematical Sciences of Anhui University since 2007. Since April 2012, he has been the Associate Professor with the School of Mathematical Sciences, Anhui University of China.

His research interests include algebraic coding theory and cryptography. He is the author of more than 60 journal articles and of one book. He has been the Area Editor of Journal of Algebra Combinatorics Discrete Structure and Application since 2014.

He has held visiting positions in School of Physical \& Mathematical Sciences, Nanyang Technological University, Singapore, from August 2012 to August 2013, Telecom ParisTech, Paris, from July 2016 to August 2016, Chern Institute of Mathematics in Nankai University since 2013. \\

Yue Guan, she has been an undergraduate student at the School of Mathematical Sciences of Anhui University since 2013. She won Honorable Mention of Mathematical Contest in Modeling Certificate of Achievement in 2016 and gained scholarships several times during school years. She will become a graduate student at School of Mathematical Sciences of Anhui University in 2017. Her research direction includes coding theory and cryptography. \\

Patrick Sol\'{e}  received the Ing\'{e}nieur and Docteur-Ing\'{e}nieur degrees both from Ecole Nationale Sup\'{e}rieure des T\'{e}l\'{e}communications, Paris, France, in 1984 and 1987, respectively, and the habilitation ид diriger des recherches from Universit\'{e} de Nice-Sophia Antipolis, Sophia Antipolis, France, in 1993.

He has held visiting positions in Syracuse University, Syracuse, NY, from 1987 to 1989, Macquarie University, Sydney, Australia, from 1994 to 1996, and Lille University, Lille, France, from 1999 to 2000.

Since 1989, he has been a permanent member of the CNRS and became Directeur de Recherche in 1996. He is currently member of the CNRS lab LAGA, from University of Paris 13.

His research interests include coding theory (codes over rings, quasi-cyclic codes), interconnection networks (graph spectra, expanders), vector quantization (lattices), and cryptography (boolean functions, pseudo random sequences).

He is the author of more than 150 journal articles and of three books.

He was the associate editor of the IEEE Information Theory Transactions from 1996 till 1999. He has been associate editor of Advances in Mathematics of Communication since 2007.

\section*{Acknowledgment}

It is the authors' pleasure to thank the anonymous referees for their helpful comments which led to improvements of the paper.

\ifCLASSOPTIONcaptionsoff
  \newpage
\fi

\end{document}